\documentclass [11pt,a4paper] {article}

\usepackage[cp1252]{inputenc}
\usepackage[english]{babel}

\usepackage{amssymb}
\usepackage{amsmath}
\usepackage{amsfonts,amssymb}

\usepackage[dvips]{graphicx}

\DeclareMathAlphabet{\mathpzc}{OT1}{pzc}{m}{it}

\begin{document}

\title{Correspondence Principle as Equivalence of Categories}

\author{Arkady Bolotin\footnote{$Email: arkadyv@bgu.ac.il$} \\ \textit{Ben-Gurion University of the Negev, Beersheba (Israel)}}

\maketitle

\begin{abstract}\noindent If quantum mechanics were to be applicable to macroscopic objects, classical mechanics would have to be a limiting case of quantum mechanics. Then the category \textbf{Set} that packages classical mechanics has to be in some sense a ‘limiting case’ of the category \textbf{Hilb} packaging quantum mechanics. Following from this assumption, quantum-classical correspondence can be considered as a mapping of the category \textbf{Hilb} to the category \textbf{Set}, i.e., a functor from \textbf{Hilb} to \textbf{Set}, taking place in the macroscopic limit. As a procedure, which takes us from an object of the category \textbf{Hilb} (i.e., a Hilbert space) in the macroscopic limit to an object of the category \textbf{Set} (i.e., a set of values that describe the configuration of a system), this functor must take a finite number of steps in order to make the equivalence of \textbf{Hilb} and \textbf{Set} verifiable. However, as it is shown in the present paper, such a constructivist requirement cannot be met in at least one case of an Ising model of a spin glass. This could mean that it is impossible to demonstrate the emergence of classicality totally from the formalism of standard quantum mechanics.\\

\end{abstract}

\section{Introduction}

\noindent According to Dirac's famous \textit{Principles of quantum mechanics} \cite{Dirac}, classical mechanics is a limiting case of more general quantum mechanics. This implies that between classical and quantum mechanics there is a correspondence principle, which demands that under certain circumstances quantum laws of motion must become equivalent to classical ones and correspondingly quantum calculations must agree with classical calculations.\\

\noindent In more general sense, the term "correspondence principle" means the reduction of one model of the reality (say, A) to another model of the reality (say, B) in appropriate circumstances. This requires that the model A explain all the phenomena under circumstances for which the model B is known to be valid, the "correspondence limit". In terms of the mathematical formalism, this means that the formal system of the model A (i.e., the set of axioms and the set of inference rules on which the model A is built on) should transform into the formal system of the model B in the correspondence limit.\\

\noindent It is reasonable to believe that every model of the reality (such as a physical theory) can be packaged in \textit{the concept of category}. In view of that, the correspondence principle could stand for an equivalence existing under certain circumstances between categories packaging the models A and B even if under all other conditions such categories stayed different.\\

\noindent For example, quantum mechanics can be packaged by the category \textbf{Hilb}, where the objects are (complex) Hilbert spaces $\mathcal{H}$ whose elements are state vectors $\left|\!\left.{\Psi} \!\right.\right\rangle$ and the morphisms are linear operators $\hat{L}: \left|\!\left.{\Psi} \!\right.\right\rangle \rightarrow \left|\!\left.{\Phi} \!\right.\right\rangle$ between vectors $\left|\!\left.{\Psi} \!\right.\right\rangle$ and $\left|\!\left.{\Phi} \!\right.\right\rangle$ of $\mathcal{H}$ \cite{Heunen,Abramsky}. Classical mechanics, on the other hand, can be packaged by the  completely different category \textbf{Set} whose objects are sets of values $X$ describing the configurations of a system and the morphisms are usual (real-valued) functions\footnote{Those functions can be partial, injective or surjective. In each case, the category \textbf{Set} would be, of course, different; however, such a difference is of no importance here.} $f: X \rightarrow Y$ from one set, $X$, to another, $Y$ \cite{Baez}. But since Bohr's correspondence principle demands that quantum mechanics and classical physics give the same answer when the systems become macroscopically large \cite{Bohr}, it seems like a logical next step to assume that in \textit{the macroscopic limit}\footnote{It is a limit for a large number $N \rightarrow \infty$ of system's constituent particles, such as atoms or molecules, where the size of the system is taken to grow in proportion with $N$. In this limit, which is a special case of the limit $\hbar \rightarrow 0$, microscopic systems are considered finite, whereas macroscopic systems are infinite; see, for example, the discussion in \cite{Landsman}.} the category \textbf{Hilb} becomes equivalent to the category \textbf{Set}. This assumption means that at the limit $N \rightarrow \infty$ there is such a “mapping” between the categories \textbf{Hilb} and \textbf{Set} (denoted as a functor $F_{\infty}$ from \textbf{Hilb} to \textbf{Set})\\

\begin{equation} \label{Functor} 
   \begin{array}{cl}
     N \rightarrow \infty:
      &
      F_{\infty}: \mathbf{Hilb}\rightarrow \mathbf{Set}
      \;\;\;\;   
   \end{array}  
\end{equation}
\smallskip

\noindent that is simultaneously full (surjective), faithful (injective) and dense (essentially surjective).\\

\noindent Specifically, in the macroscopic limit the functor $F_{\infty}$ would associate to any object $\mathcal{H}$ of the category \textbf{Hilb} the matching well-defined set of the category $\mathbf{Set}$, i.e.\\

\begin{equation} \label{Object} 
     \forall \mathcal{H} \in \mathbf{Hilb}
      :\;\;
      F_{\infty}\! \left(\mathcal{H} \right) \in \mathbf{Set}
      \;\;\;\;   ,
\end{equation}
\smallskip

\noindent and to any linear operator $\hat{L}: \left|\!\left.{\Psi} \!\right.\right\rangle \rightarrow \left|\!\left.{\Phi} \!\right.\right\rangle$ of \textbf{Hilb} the matching real-valued function of $\mathbf{Set}$, i.e.\\

\begin{equation} \label{Operator} 
     \forall \hat{L}: \left|\!\left.{\Psi} \!\right.\right\rangle \rightarrow \left|\!\left.{\Phi} \!\right.\right\rangle \in \mathbf{Hilb}
      :\;\;
             F_{\infty} \! \left(\hat{L} \right): 
                                     F_{\infty} \!\left(\left|\!\left.{\Psi} \!\right.\right\rangle \right) 
                                     \rightarrow 
                                     F_{\infty} \!\left(\left|\!\left.{\Phi} \!\right.\right\rangle \right)  \in \mathbf{Set}
      \;\;\;\;   .
\end{equation}
\smallskip

\noindent In addition, it is logical to demand that the functor $F_{\infty}$ would be naturally isomorphic to any other functor $G_{\infty}$, which also maps the category \textbf{Hilb} to the category \textbf{Set} in the macroscopic limit $N \rightarrow \infty$.\\

\noindent Since the functor $F_{\infty}$ is a mathematical construction that has an exact characterization, we will assume that the list of the rules (\ref{Object}) and (\ref{Operator}) as descriptions of allowable steps of the associating process is given. In other words, we will assume that the functor $F_{\infty}$ exists only if there is a way (i.e., an effective procedure) \textit{to construct it}.\\

\noindent In contrast, suppose that the existence of the functor $F_{\infty}$ is proved but no explicit rules how to map every object of \textbf{Hilb} to an object of \textbf{Set} and every morphism in \textbf{Hilb} to a morphism in \textbf{Set} is provided. In that case, the potential realizability of the functor $F_{\infty}$ and thus the verifiability of Bohr's correspondence principle cannot be guaranteed.\\

\noindent What follows from the assumption of the functor $F_{\infty}$ \textit{constructivism} is that the computability of the set $F_{\infty}\! \left(\mathcal{H} \right)$, that is, the existence of an algorithm able to identify the exact members of the set $F_{\infty}\! \left(\mathcal{H} \right)$, is a requisite for the existence of the verifiable equivalence of the categories \textbf{Hilb} and \textbf{Set} in the macroscopic limit $N \rightarrow \infty$.\\

\noindent Even though such a requisite may seem to be rather benign\footnote{See, for example, the paper regarding the contraction of groups and their representations \cite{Wigner}, the papers on phase-space representation of state vectors \cite{Ban, Curtright}, and also texts on the quantum theory of infinite systems such as \cite{Araki, Sewell}.}, in the present paper it will be shown that for at least one quantum Ising model of a spin glass, the set $F_{\infty}\! \left(\mathcal{H} \right)$ is not computable (that is, not effectively possible). This could mean that it is impossible to truly represent classical mechanics in quantum mechanics and hence to actually demonstrate the emergence of classicality purely from the formalism of standard quantum mechanics.\\

\section{The macroscopic Ising model of the spin glass}

\noindent Let us consider the quadratic function $H\!(\!\boldsymbol{S}_k\!)$ of a set of $N$ discrete variables (“spins”) $S_{ik}=\pm1$\\

\begin{equation} \label{Quadratic_function} 
     H\!(\!\boldsymbol{S}_{\!k}\!)
      =
      \left(
      \sum_{i=1}^{N}q_i S_{ik}
      \right)^2
      \;\;\;\;   ,
\end{equation}
\smallskip

\noindent where $q_i$ are positive numbers of arbitrary size or -- after division by the maximal number $\max_{i} {q_i}$ -- unlimited precision, and $\boldsymbol{S}_k=(S_{1k},S_{2k},\dotsc, S_{Nk})$ denotes the $k^{\mathrm{th}}$  configuration of variables $S_{ik}$. It is easy to notice that this function can be written as the Hamiltonian $H_{\mathrm{Ising}}(\!\boldsymbol{S}_k\!)$ of the infinite-range antiferromagnetic Ising model of a spin glass\\

\begin{equation} \label{Ising_Hamiltonian} 
     H_{\mathrm{Ising}} (\!\boldsymbol{S}_{\!k}\!)
      =
      \sum_{n_1+n_2+\cdots+n_N=2}
            \! \binom{2}{n_1,n_2,\dots,n_N}
                    q_1^{n_1}q_2^{n_2}\cdots q_N^{n_N}
                          S_{1k}^{n_1}S_{2k}^{n_2} \cdots S_{Nk}^{n_N}
      \;\;\;\;   ,
\end{equation}
\smallskip

\noindent where all possible pairs of spins $S_{ik}S_{jk}$ have interactions (for reviews on spin glasses, see, for example, \cite{Bryngelson, Fischer, Nishimori}).\\

\noindent Let us find the lowest possible macroscopic energy of this Ising model. According to the general prescription of statistical mechanics \cite{Landau}, the macroscopic energy of the Ising model $\langle E \rangle$ (i.e., the average energy of the model in the macroscopic limit $N \rightarrow \infty$) is given by the formula\\

\begin{equation} \label{Macroscopic_energy} 
     \langle E \rangle
      =
      -\frac{\partial}{\partial{\beta}} \ln{Z}
      \;\;\;\;   ,
\end{equation}
\smallskip

\noindent where $\beta$ is the inverse thermodynamic temperature $T^{-1}$ (assuming that Boltzmann's constant $k_B$ is unity) and $Z$ is the partition function defined over all possible spin configurations $\boldsymbol{S}_k$\\

\begin{equation} \label{Partition} 
     Z
      =
      \sum_{k}^{2^N} e^{-\beta H\!(\boldsymbol{S}_{\!k})}
      \;\;\;\;   .
\end{equation}
\smallskip

\noindent Let us consider the ratio of probabilities $W\!(\!\boldsymbol{S}_k\!)$ and $W\!(\!\boldsymbol{S}_m\!)$ of being in the state with configuration $\boldsymbol{S}_k$ or $\boldsymbol{S}_m$ defined by the Boltzmann factor\\

\begin{equation} \label{Boltzmann_factor} 
     \frac{W\!(\!\boldsymbol{S}_{\!k}\!)}{W\!(\!\boldsymbol{S}_{\!m}\!)}
      =
     e^{-\beta \big(
                                   H\!(\boldsymbol{S}_{\!k}) - H\!(\boldsymbol{S}_{\!m} )
                       \big)
          }
      \;\;\;\;   .
\end{equation}
\smallskip

\noindent We notice then that at low temperatures $\beta \gg 1$ the Boltzmann factor (\ref{Boltzmann_factor}) indicates that only configurations minimizing the quadratic function (\ref{Quadratic_function}) will contribute to the partition function $Z$, while at high temperatures $\beta \ll 1$ the configurations with various values (including high ones) of the function (\ref{Quadratic_function}) will appear in $Z$ with similar probabilities.\\

\noindent From here, one can deduce that as a function of the temperature $\beta$ the macroscopic energy $\langle E \rangle$ approaches its lowest value $\min\!{\langle E \rangle}$ as $\beta$ approaches infinity, namely,\\

\begin{equation} \label{Min_E} 
     \min\!{\langle E \rangle}
      =
      -\lim_{T \rightarrow 0} {T\ln{Z}}
      \;\;\;\;   ,
\end{equation}
\smallskip

\noindent where the limit $T \rightarrow 0$ makes certain that only configurations $\boldsymbol{S}_k$ that secure the global minimum of the function (\ref{Quadratic_function}) will be enclosed in the partition $Z$. And since the global minimum of this function exists by construction, $\min\!{\langle E \rangle}$ exists as well. It is essential to note here that the existence of $\min\!{\langle E \rangle}$ is assured \textit{apart from any consideration of the possibility of its construction} (i.e., calculation).\\

\section{The quantum Ising model of the spin glass}

\noindent Next, let us consider the quantum version of the Ising model (\ref{Quadratic_function})\\

\begin{equation} \label{Quantum_version} 
     H\!\left(\!  \sigma_{1}^{z},\sigma_{2}^{z}, \dotsc, \sigma_{N}^{z}  \!\right)
      =
      \left(
      \sum_{i=1}^{N}q_i \sigma_{i}^{z}
      \right)^2
      \;\;\;\;   ,
\end{equation}
\smallskip

\noindent where each discrete variable $S_{ik}$ is replaced by the Pauli matrix $\sigma_{i}^{z}$ acting on the $i^{\mathrm{th}}$ qubit labeled by $\left|\!\left.{z_{ik}} \!\right.\right\rangle$ with $z_{ik} \in\{-1,+1\}$ such that $S_{ik}=+1$ corresponds to $\left|\!\left.{z_{ik}=+1} \!\right.\right\rangle$ (i.e., the $i^{\mathrm{th}}$ quantum spin being up in the $z$-direction) and $S_{ik}=-1$ corresponds to $\left|\!\left.{z_{ik}=-1} \!\right.\right\rangle$ (i.e., the $i^{\mathrm{th}}$ quantum spin down in the $z$-direction). In this way, the Hilbert space of the considered quantum model would be spanned by the $2^N$ basis vectors $\left|\!\left.{z_{1k}} \!\right.\right\rangle, \left|\!\left.{z_{2k}} \!\right.\right\rangle, \dotsc, \left|\!\left.{z_{Nk}} \!\right.\right\rangle$.\\

\noindent Let $\left|\!\left.{\psi_{k}} \!\right.\right\rangle = \left|\!\left.{z_{1k}} \!\right.\right\rangle \left|\!\left.{z_{2k}} \!\right.\right\rangle \cdots \left|\!\left.{z_{Nk}} \!\right.\right\rangle$ denote the $k^{\mathrm{th}}$ spin configuration of the quantum Ising model, i.e., the  $k^{\mathrm{th}}$ eigenstate of the Hamiltonian (\ref{Quantum_version}). Also, let  $A_k$ be the subset of the index set $I=\{1,2,\dotsc,N\}$ which identify the spins in the configuration $\left|\!\left.{\psi_{k}} \!\right.\right\rangle$  being up in the z-direction:

\begin{equation} \label{Subset_A} 
     A_k
      =
      \big\{
      i \in {I} \,\big |\,
                                   \left|\!\left.{z_{ik}=+1} \!\right.\right\rangle  \!
      \big\}
      \;\;\;\;   .
\end{equation}
\smallskip

\noindent Then, as it is readily apparent, the eigenvalues $E_k$ of the Hamiltonian (\ref{Quantum_version}) can be presented as the outputs of the function $E(A_k)$ defined by the set of the subsets $A_k$\\

\begin{equation} \label{Discrepancy} 
     E_k
      =
     E\!\left(A_k\right)
      =
      \Big(
      \sum_{i \in A_k}q_i \; - \sum_{i \in I\setminus A_k}\! q_i
      \Big)^2
      \;\;\;\;   .
\end{equation}
\smallskip

\noindent Accordingly, the problem of finding the ground energy $E_{\mathrm{ground}}$  of the quantum Ising model (\ref{Quantum_version})\\

\begin{equation} \label{SE} 
      H\!\left(\!  \sigma_{1}^{z},\sigma_{2}^{z}, \dotsc, \sigma_{N}^{z}  \!\right)
                                                                                                                                               \left|\!\left.{\psi_{\mathrm{ground}}} \!\right.\right\rangle
      =
      E_{\mathrm{ground}}
                                                                                                                                               \left|\!\left.{\psi_{\mathrm{ground}}} \!\right.\right\rangle
      \;\;\;\;    
\end{equation}
\smallskip

\noindent would be equivalent to the problem of finding the subsets $A_k=A_{\mathrm{ground}}$ that minimize the function $E (A_k)$, i.e.,\\

\begin{equation} \label{Min_discrepancy} 
     E_{\mathrm{ground}}
      =
     E\!\left(A_{\mathrm{ground}}\right)
      =
      \min {\Big\{E\!\left(A_k\right)\!\Big\}_{k=1}^{2^N}}
      \;\;\;\;   ,
\end{equation}
\smallskip

\noindent the latter problem is called the \textit{two-way number partitioning problem}, $\mathrm{N}_{\mathrm{PP}}$.\\

\noindent As it is known, the $\mathrm{N}_{\mathrm{PP}}$ is one of the classical NP-hard problems of combinatorial optimization \cite{Jones, Mertens01}. In particular, this means that for all numbers $q_i$ bounded from above by the value of $2^{\alpha N}$, where $\alpha >0$, to exactly solve the equation (\ref{SE}) might take an exponential in $N$ amount of time for all $\alpha$ \cite{Mertens03}. More explicitly, in the worst case it would take $O(2^{N/2})$ amount of time and $O(2^{N/4})$ amount of space, according to the best known exact algorithms for solving the  $\mathrm{N}_{\mathrm{PP}}$ \cite{Boettcher, Ferreira, Horowitz, Schroeppel}.\\

\section{Correspondence principle between the quantum and classical Ising models of the spin glass}

\noindent The important question that needs to be asked now is this: Can the verifiable (computable) equivalence between the quantum and classic Ising models of the spin glass exist in the macroscopic limit $N \rightarrow \infty$?\\

\noindent Recall that the verifiable equivalence between these two models consists of the functor $F_{\infty}$ that can in an effectively calculable way (i.e., in a finite amount of time) map the Hilbert spaces $\mathcal{H}$ of the quantum Ising model at the limit $N \rightarrow \infty$ to the configurations $\boldsymbol{S}_k$ of classic Ising model at the same limit. This would imply that the functor $F_{\infty}$ associates to the eigenspace $\mathcal{E} (E_{\mathrm{ground}}) \in \mathbf{Hilb}$ of the ground energy $E_{\mathrm{ground}}$  of the Hamiltonian $H(\! \sigma_{1}^{z},\sigma_{2}^{z}, \dotsc, \sigma_{N}^{z} \!)$ in the limit $N \rightarrow \infty$ the classical configurations $F_{\infty}\! \left(\mathcal{E} (E_{\mathrm{ground}})\right) \in \mathbf{Set}$ corresponding to the macroscopic state of the spin glass with lowest possible energy $\min\!{\langle E \rangle}$.\\

\noindent Because defining the property of being the eigenspace $\mathcal{E} (E_{\mathrm{ground}})$ in the limit $N \rightarrow \infty$ involves quantification over the uncountably infinite domain\\

\begin{equation} \label{E_k_limit} 
     \lim_{N \rightarrow \infty} \mathcal{E} \left(E_{\mathrm{ground}}\right)
     =
     \mathrm{Null} \bigg(
                                            H - \min_k {\Big\{E\!\left(A_k\right)\!\Big\}_{k=1}^{2^{\infty}} }\; \hat{I}
                                 \bigg)
      \;\;\;\;    
\end{equation}
\smallskip

\noindent (where $\mathrm{Null}(\cdot )$ stands for the null space corresponding to the eigenspace $\mathcal{E} (E_{\mathrm{ground}})$ at the macroscopic limit and $\hat{I}$ is the identity operator), identifying the ground energy $E_{\mathrm{ground}} \ne 0$ of the quantum Ising model in this limit by the best known exact algorithms for solving the $\mathrm{N}_{\mathrm{PP}}$ would require checking the inequality $E_{\mathrm{ground}} \le E\!\left(A_k\right)$ infinitely many times\footnote{It means that the procedure for finding the global minimum $\mathcal{E} (E_{\mathrm{ground}})$  would be undecidable in the sense of Church-Turing thesis of computability \cite{Ben_Amram}.}.\\

\noindent Thus, from the constructivist point of view the functor $F_{\infty}$ cannot exist. Consequently, the equivalence of quantum and classical Ising models of the spin glass in the macroscopic limit cannot be actually demonstrated. And so, providing the presumed universal nature of the correspondence principle, one can infer from here that the assumption of the verifiable equivalence of the categories \textbf{Hilb} and \textbf{Set} in the macroscopic limit $N \rightarrow \infty$ cannot be correct.\\

\section{Concluding remarks}

\noindent One obvious objection can be brought up here: Strictly speaking, the domain of eigenvalues $E(A_k)$ is infinite only in the hypothetical case of an infinitely large spin glass (containing an infinite number $N$ of “spins” $S_{ik}=\pm 1$). But since the assumption of infinite sizes is never realized in nature \cite{Bell}, one can suppose that the category $\mathbf{Hilb}$ might be equivalent to the category $\mathbf{Set}$ in the more ‘realistic’ macroscopic limit $N \rightarrow N_{\!A}$, where $N_{\!A} \sim 10^{24}$ is Avogadro's number.\\

\noindent However, inasmuch as the Hamiltonian of the quantum Ising model is NP-hard, replacing actual infinity by the ‘realistic’ one, $N_{\!A}$, would make no difference. Indeed, unless the $\mathrm{N}_{\mathrm{PP}}$ can be proved NP-easy and thus solved by a deterministic Turing machine in the number of steps upper bounded by a polynomial expression in the size of $N_{\!A}$, finding the ground energy $E_{\mathrm{ground}}$ of the quantum Ising Hamiltonian (\ref{Min_discrepancy}) in the limit $N \rightarrow N_{\!A}$ would require searching through all $2^{N_{\!A}}$ eigenvalues $E_k$ (or no less than $2^{N_{\!A}/2}$ of them). Clearly, this would take an amount of time that – in every practical sense – does not differ much from actual infinity.  As a result, the equivalence of quantum and classical Ising models of the spin glass in the ‘realistic’ macroscopic limit $N \rightarrow N_{\!A}$ cannot be considered verifiable as well.\\

\noindent The demonstrated above constructivist incomparability of the categories $\mathbf{Hilb}$ and $\mathbf{Set}$ indicates that it is impossible to construct (to compute) the classical motion of macroscopic bodies within the formalism of standard quantum mechanics or, in other words, to actually verify that classical mechanics is a limiting case of quantum mechanics\footnote{ Of course, it is possible to object that saying that the laws of classical mechanics follow from the laws of quantum mechanics only as \textit{an approximation} at the limit of large systems; however, such an argument undermines the exact meaning of the equivalence principle.}.\\

\noindent Such a conclusion is in line with the argument, which asserts that the problem of the emergence of classical mechanics from quantum mechanics is still open \cite{Allori}. As this argument goes, if classicality is associated with the formation and preservation of narrow wave packets, one should admit that those wave packets typically spread and there is a definite time, after which the classical approximation will break down, and it can easily be shown that interactions will typically generate very spread out wave functions, even for massive bodies.\\

\end{document}